\documentclass[twocolumn]{aastex6}
\usepackage{bm} 
\usepackage{color}

\usepackage{graphicx}
\usepackage[subfigure]{graphfig}
\usepackage{amsmath}
\usepackage[figuresright]{rotating}

\newcommand{\lya}{\hbox{Ly$\alpha$}}
\newcommand{\hb}{\hbox{H$\beta$}}
\newcommand{\mgii}{\hbox{Mg\,{\sc ii}}}
\newcommand{\nev}{[\hbox{Ne\,{\sc v}}]}
\newcommand{\oii}{[\hbox{O\,{\sc ii}}]}

\newcommand{\neiii}{[\hbox{Ne\,{\sc iii}}]}
\newcommand{\oiii}{[\hbox{O\,{\sc iii}}]}
\newcommand{\feii}{\hbox{Fe\,{\sc ii}}}
\newcommand{\feiii}{\hbox{Fe\,{\sc iii}}}

\newcommand{\civ}{\hbox{C\,{\sc iv}}}
\newcommand{\hei}{\hbox{He\,{\sc i}}}

\newcommand{\hf}{\hbox{H$\zeta$}}
\newcommand{\he}{\hbox{H$\epsilon$}}

\fullcollaborationName{The Friends of AASTeX Collaboration}

\begin{document}


\title{relation between the variations in the $\mgii\ \lambda2798$ emission-line and 3000 \AA \ continuum}



\author{Dongchun Zhu\altaffilmark{1,2}, Mouyuan Sun\altaffilmark{1,2}, Tinggui Wang\altaffilmark{1,2}}
\affil{$^1$CAS Key Laboratory for Research in Galaxies and Cosmology, Department of Astronomy, University of Science and Technology of China, Hefei 230026, China; zhudc@mail.ustc.edu.cn\\$^2$School of Astronomy and Space Science, University of Science and Technology of China, Hefei 230026, China}








\begin{abstract}
We investigate the relationship between the $\mgii \ \lambda2798$ emission-line and the 3000 \AA \ continuum variations using a sample of 68 intermediate-redshift ($z\sim$ 0.65$-$1.50) broad-line quasars spanning a bolometric luminosity range of 44.49 erg s$^{-1} \leq \rm{log}$$L_{\rm{bol}} \leq 46.31$ erg s$^{-1}$ (Eddington ratio from $\sim$ 0.026 to 0.862). This sample is constructed from SDSS-DR7Q and BOSS-DR12Q, each with at least 2 spectroscopic epochs in SDSS-I/II/III surveys. Additionally, we adopt the following signal-to-noise ratio (S/N) selection criteria: a) for $\mgii$ and the 3000 \AA \ continuum, S/N $\geq$ 10; b) for narrow lines, S/N $\geq$ 5. All our quasar spectra are recalibrated based on the assumption of constant narrow emission-line fluxes. In an analysis of spectrum-to-spectrum variations, we find a fairly close correlation (Spearman $\rho = 0.593$) between the variations in broad $\mgii$ and in the continuum. This is consistent with the idea that $\mgii$ is varying in response to the continuum emission variations. Adopting the modified weighted least squares regression method, we statistically constrain the slopes (i.e., the responsivity $\alpha$ of the broad $\mgii$) between the variations in both components for the sources in different luminosity bins after eliminating intrinsic biases introduced by the rescaling process itself. It is shown that the responsivity is quite small (average $\bar{\alpha} \approx$ 0.464) and anti-correlates with the quasar luminosity. Our results indicate that high signal-to-noise flux measurements are required to robustly detect the intrinsic variability and the time lag of $\mgii$ line.
\end{abstract}




\section{Introduction} \label{sec:intro}

It is now widely accepted that quasars are powered by accretion of material onto supermassive black holes (SMBHs). The continuum emission and the broad emission lines (BELs) often show aperiodic variations (e.g., \citealt{FitchEtAl1967}; \citealt{AndrillatEtAl1968}). Theoretically, the BEL fluxes are supposed to vary in response to the variations of the ionizing continuum with a lag of about light travelling time. Hence, via the cross correlation analysis of the BELs and the continuum, we are able to constrain the geometry of the spatially unresolved broad emission-line region (BLR) in active galactic nucleus (AGN). Notably, with the light-travel time delay, the distance of the BLR to the ionizing source is directly determined (e.g., \citealt{Peterson1993}). By an attempt to model that the BLR is virialized, the central SMBH mass then can be estimated (e.g., \citealt{WandelEtAl1999}).

So far emission-line reverberation mapping (RM; e.g., \citealt{BlandfordEtAl1982}) experiments have succeeded in measuring emission line lags in $\geq$ 60 AGNs; (e.g., \citealt{PetersonEtAl1998, PetersonEtAl2002, PetersonEtAl2004}; \citealt{WandelEtAl1999};
\citealt{KaspiEtAl2000, KaspiEtAl2005}; \citealt{VestergaardEtAl2006}; \citealt{BentzEtAl2009, BentzEtAl2013}; \citealt{DenneyEtAl2010}; \citealt{BarthEtAl2011, BarthEtAl2011a}; \citealt{GrierEtAl2012}; \citealt{Hu2015}; \citealt{GoadEtAl2016}; \citealt{JiangEtAl2016}; \citealt{ShenEtAl2016}). It is revealed that, the BLR size as measured for a particular emission line such as $\hb \ \lambda4861$, is closely related to the AGN luminosity in the approximate form $R \propto L^{1/2}$ (the $R$-$L$ relation; e.g., \citealt{KaspiEtAl2000}; \citealt{BentzEtAl2006};  \citealt{ShenEtAl2012}). This relation offers the possibility of taking advantage of single-epoch (SE) spectra to determine the SMBH masses (e.g., \citealt{Vestergaard2002}; \citealt{McLureEtAl2002}; \citealt{VestergaardEtAl2006}). Over the past decade, several editions of these estimations have been developed (see, e.g., \citealt{McGillEtAl2008}; \citealt{WangEtAl2009a}). Resulted from the economical efficiency and operability, the SE virial SMBH mass estimation is a sort of praticable method on the determination of AGN SMBH masses compared to RM technique (e.g., \citealt{WooEtAl2002}; \citealt{McLureEtAl2004}).

RM studies, as they are known, have been traditionally performed mostly on low-luminosity AGNs at low redshift ($z <$ 0.3) using the $\hb$ emission-line to measure SMBH masses. For AGNs at redshifts beyond 1, rest-frame ultraviolet (UV) BELs are required, such as $\mgii$, a crucial emission line of RM interest that can be presented in quasar spectra having redshifts between 0.3 and 2. However, RM results of $\mgii$ line (i.e., reliable detection of $\mgii$ lag) are quite scarce. This is initially interpreted that the $\mgii$ emission-line varies more slowly in response to continuum changes than $\hb$ emission line, suggesting that the $\mgii$-emitting region may have larger-scale structure than that of $\hb$ (e.g., \citealt{CorbettEtAl2003}).

The relationship between the BEL flux ($F_{line}$) and the continuum flux ($F_{cont}$) within an individual source is often expressed by $F_{line}$ $\propto$ $F_{cont}^{\alpha}$, where $\alpha$ is traditionally measured from emission line and continuum flux light curves from AGN monitoring campaigns. This is related to the so-called intrinsic \textquotedblleft Baldwin Effect" (see, e.g., \citealt{KinneyEtAl1990}; \citealt{PoggeEtAl1992}; \citealt{GoadEtAl2004}; \citealt{KoristaEtAl2004}). 
In fact, $\alpha$ is commonly referred to as the response of the BEL to variations in the ionizing continuum flux. Formally, we can parameterize the correlation between the variations in the $\mgii\ \lambda2798$ line ($\rm{dlog}$$F_{line}$) and in the 3000 \AA \ continuum ($\rm{dlog}$$F_{cont}$) with a simple linear function of the form $\rm{dlog}$$F_{line} \propto$ $\rm{dlog}$$F_{cont}$. Given that the ratio in magnitude changes ought to be equivalent to $\alpha$, the value of $\mgii$ responsivity can be calculated by analysing spectroscopic monitoring data.

When determining the emission line responsivity parameter $\alpha$ = $\rm{dlog}$$F_{line}$/$\rm{dlog}$$F_{cont}$, it is of great importance to ensure that the emission-line flux is referenced to the correct (in time) continuum value (e.g., \citealt{GoadEtAl2004}; \citealt{GoadEtAl2014}). Generally, this parameter is determined from temporally well-sampled continuum and emission line light curves, and the correct reference continuum is determined by shifting the emission line light curve backward in time by the emission line lag. In practice, the $\mgii$ lag is not well-constrained in prior RM campaigns (e.g., \citealt{ClavelEtAl1991a}; \citealt{CackettEtAl2015}, but see, \citealt{ReichertEtAl1994}; \citealt{MetzrothEtAl2006}), and few robust measurements of its responsivity have been measured. Instead of temporally well-sampled light curves of a single AGN, in this study, we use 1210 data pairs of spectroscopic observations of 68 AGNs to {\em statistically} estimate the responsivity of the broad $\mgii$ emission line. Here, we posit that our ignorance of the emission line lag corrections averages out statistically, and an ensemble responsivity of $\mgii$ may be determined from many pairs of measurements from an amount of AGNs.

Reliable flux calibration is of significance to accurately determine the observed emission-line and continuum flux. The Sloan Digital Sky Survey (SDSSS; \citealt{YorkEtAl2000}) spectroscopy is routinely calibrated using a series of standard stars, particularly main sequence F stars. 
For a single observation, it is assumed that the uncertainty of SDSS-I/II spectroscopic data is $\sim$ 0.04 mag (\citealt{Adelman-McCarthyEtAl2008}). With the smaller fibers, SDSS-III BOSS (e.g., \citealt{MargalaEtAl2015}; \citealt{HarrisEtAl2016}) spectroscopy is usually not as accurate as that of SDSS-I/II. In this work, we assume that the fluxes of narrow emission line have no variations during the spectroscopic monitoring due to the much large narrow-line region (NLR). Therefore, we attempt to use narrow-line fluxes to recalibrate SDSS quasar spectra. All the flux variations are measured using ground-based optical monitoring data from the observed flux of emission-lines and continuum in any two epochs during SDSS-I/II/III surveys. 

The structure of this paper is as follows. In Section \ref{sec:style} we describe our quasar sample selection. In Section \ref{sec:sm} we introduce the details of our spectral measurements. We derive the correlation between the variations of $\mgii \ \lambda$2798 and of the 3000 \AA \ continuum in the SDSS quasars in Section \ref{sec:results}. We discuss the related results in Section \ref{sec:dis}, and a summary of our conclusions in Section \ref{sec:con}. Throughout this paper, we adopt a flat cosmology with $\Omega_m$ = 0.3, $\Omega_{\Lambda}$ = 0.7, and $H_0$ = 70 km s$^{-1}$ Mpc$^{-1}$, and use magnitude (rather than flux or luminosity) differences to characterize variations. Unless otherwise specified, the reported wavelengths (taken from \citealt{BerkEtAl2001}) and timescales are in the quasar rest-frame.

\section{THE Sample} \label{sec:style}

In this work, we use the quasar data from the compilation of the SDSS Data Release 7 Quasar catalog (DR7Q; e.g., \citealt{SchneiderEtAl2010}; \citealt{ShenEtAl2011}) and Data Release 12 Quasar catalog (DR12Q; e.g., \citealt{ParisEtAl2014}; \citealt{ParisEtAl2017}). 
All the spectra were taken by the Apache Point 2.5 m wide-field telescope (\citealt{GunnEtAl2006}) during SDSS-I/II/III surveys (2000-2014). Each spectrum is stored in vacuum wavelength with a resolution of $R \sim 1500-2500$.

Our parent sample was compiled from the following 2 sub-samples: the DR7Q consisted of 105,783 objects that are brighter than $M_i = -22.0$, and the DR12Q including 297,301 quasars. There are 7,063 quasars from DR7Q and 28,105 quasars from DR12Q, each with multiple ($\geq$ 2) spectroscopic epochs, respectively. After confirming the quasar as a point-source in the SDSS image and rejecting the epoch with low-quality spectrum, we selected a sample of 2,374 quasars with $\mgii$ broad-line by requiring $0.65 \leq z \leq 1.50$. This requirement ensures that broad $\mgii$, narrow-lines (e.g., $\oiii \ \lambda\lambda4960,5008$) and the 3000 \AA \ continuum region are presented in the SDSS spectra. 

We notice that additional flux deficit is confirmed in the SDSS-III Baryon Oscillation Spectroscopic Survey (BOSS; e.g., \citealt{EisensteinEtAl2011}; \citealt{BoltonEtAl2012}; \citealt{DawsonEtAl2013}; \citealt{SmeeEtAl2013a}) relative to SDSS-I/II due to the difference in flux calibration from SDSS-I/II to BOSS. To obtain an accurate and reliable measurement of the intrinsic variations in $\mgii \ \lambda2798$ and in the 3000 \AA \ continuum for each quasar, we develop an independent correction to the flux variations, which is called narrow-line flux-recalibration (see Section \ref{subsec:vc}). This requires that every quasar in our sample not only has repeated observations but also contains a minimum signal-to-noise ratio (S/N; defined as the ratio between the emission-line flux and its error) of 5 for narrow-line(s). In addition, we rejected objects with unusual emission line profiles and/or continuum shapes (i.e., BALQSOs) from our final sample. The reduced $\chi^2$ values ($\chi^2$/dof) of our best-fit model for these sources are often fairly large during emission-line fitting (see Section \ref{subsec:mu}). More details about the sample-selection criteria include the following.

\begin{enumerate}
\item Multiple ($\geq$ 2) spectroscopic epochs/observations are included for each quasar in the SDSS-I/II/III surveys.
\item Quasar is confirmed to be a point-source in the SDSS image (take example for DR7Q, {\tt\string sdss\_morpho = 0}).
\item A minimum S/N ratio of 10 for quasar spectrum covering $\mgii$ through the 3000 \AA \ continuum is preferred.
\item A redshift between 0.65 and 1.50 should be possessed for each object.
\item A minimum S/N ratio of 5 for narrow-line(s) are required in the SDSS quasar spectra.
\item Quasar with peculiar $\mgii$ emission-line and continuum property is rejected.
\end{enumerate}

Table \ref{tb1} summarizes part of the final 1210 data pairs consisting of 68 quasars that passed all the selection criteria and will be used for subsequent relation analysis of the spectroscopic variations in the $\mgii\ \lambda2798$ emission-line and in the 3000 \AA \ continuum. 

\begin{deluxetable*}{ccccccccccccccc}
\tabletypesize{\scriptsize}
\tablecaption{\label{tb1} Sample Summary}
\tablewidth{0pt}
\tablehead{
\colhead{ID} & \colhead{Object Name} & \colhead{Plate} &
\colhead{Fiber} & \colhead{MJD} & \colhead{$EW_{_{\rm{mgii}}}$} & \colhead{$\rm{log}$$L_{_{\rm{mgii}}}$} & \colhead{$\rm{log}$$L_{_{\rm{nev2}}}$} & \colhead{$\rm{log}$$L_{_{\rm{oii}}}$} & \colhead{$\rm{log}$$L_{_{\rm{neiii1}}}$} & \colhead{$\rm{log}$$L_{_{\rm{oiii2}}}$} & \colhead{$z_{_{\rm{vi}}}$} & \colhead{$\rm{log}$$L_{_{\rm{3000}}}$}  & \colhead{Catalog} & \colhead{$N_{\rm{r}}$}
\\
\colhead{(1)} & \colhead{(2)} & \colhead{(3)} & \colhead{(4)} & \colhead{(5)} & \colhead{(6)} & \colhead{(7)} & \colhead{(8)} & \colhead{(9)} & \colhead{(10)} & \colhead{(11)} & \colhead{(12)} & \colhead{(13)} & \colhead{(14)} & \colhead{(15)}
}
\startdata
   1 & J002303.15+011533.6 &0390  &562 &51816 & 37.97 & 41.26  & 41.26 & 41.20 & 41.35 &  42.28 &0.73 & 45.02 & DR07Q &2  \\
   1 & J002303.15+011533.6 &0390  &567 &51900 & 37.69 & 41.09  & 41.38 & 41.24 & 41.41 &  42.31 &0.73 & 45.09 & DR07Q &2  \\
   2 & J002303.15+011533.6 &0390  &567 &51900 & 37.69 & 41.09  & 41.38 & 41.24 & 41.41 &  42.31 &0.73 & 45.09 & DR07Q &3  \\
   2 & J002303.15+011533.6 &4300  &181 &55528 & 39.33 & 40.99  & 41.53 & 41.33 & 41.42 &  42.44 &0.73 & 45.10 & DR07Q &3  \\
   3 & J004212.19+173135.4 &6198  &799 &56211 & 59.87 & 41.05  & 41.23 & 42.08 & 41.39 &  42.47 &0.90 & 44.63 & DR12Q &3  \\
   3 & J004212.19+173135.4 &6193  &050 &56237 & 68.00 & 41.06  & 41.22 & 41.98 & 41.33 &  42.36 &0.90 & 44.61 & DR12Q &3  \\
   4 & J010033.49+002200.3 &0396  &342 &51816 & 64.69 & 42.04  & 41.86 & 42.00 & 41.97 &  42.94 &0.75 & 45.27 & DR07Q &3  \\
   4 & J010033.49+002200.3 &0693  &466 &52254 & 49.07 & 41.50  & 41.65 & 41.96 & 41.66 &  42.89 &0.75 & 45.33 & DR07Q &3  \\
   5 & J013053.43$-$095710.2 &0662  &276 &52147 & 32.17 & 41.65 & 41.88 & 41.82 & 41.71 &  42.75 &0.73 & 45.31 & DR07Q &3  \\
   5 & J013053.43$-$095710.2 &0662  &273 &52178 & 30.19 & 41.64 & 41.77 & 41.77 & 41.86 &  42.82 &0.73 & 45.32 & DR07Q &3  \\
   6 & J013053.43$-$095710.2 &0662  &276 &52147 & 32.17 & 41.65 & 41.88 & 41.82 & 41.71 &  42.75 &0.73 & 45.31 & DR07Q &3  \\
   6 & J013053.43$-$095710.2 &2878  &118 &54465 & 53.63 & 41.20 & 41.75 & 41.65 & 41.72 &  42.70 &0.73 & 45.03 & DR07Q &3  \\
  \vdots & \vdots & \vdots & \vdots & \vdots & \vdots & \vdots & \vdots & \vdots & \vdots & \vdots & \vdots & \vdots & \vdots & \vdots \\

\enddata
\tablecomments{\scriptsize Column 1: identification number of our data pairs in this paper. Column 2: SDSS object name. Column 3-5: plate, fiber, and MJD (i.e., JD-2400000) of the optical SDSS spectrum for each object. Column 6: the broad $\mgii$ EW relative to the underlying continuum at 3000 \AA. Column 7-11: luminosity measurement of narrow $\mgii$ doublet, $\nev \ \lambda3426$, $\oii \ \lambda3726,3729$ doublet, $\neiii \ \lambda3869$,  and $\oiii \ \lambda5008$ in this paper. Column 12: improved quasar redshift from catalog DR7Q and DR12Q. Column 13: the 3000 \AA \ continuum luminosity measurement for each spectroscopic epoch in our sources. Column 14: quasar catalog in which each observation is included. Column 15: the number of narrow-lines used in our final flux-recalibration.}
\end{deluxetable*}

\begin{figure*}
\includegraphics[angle=0,scale=0.975]{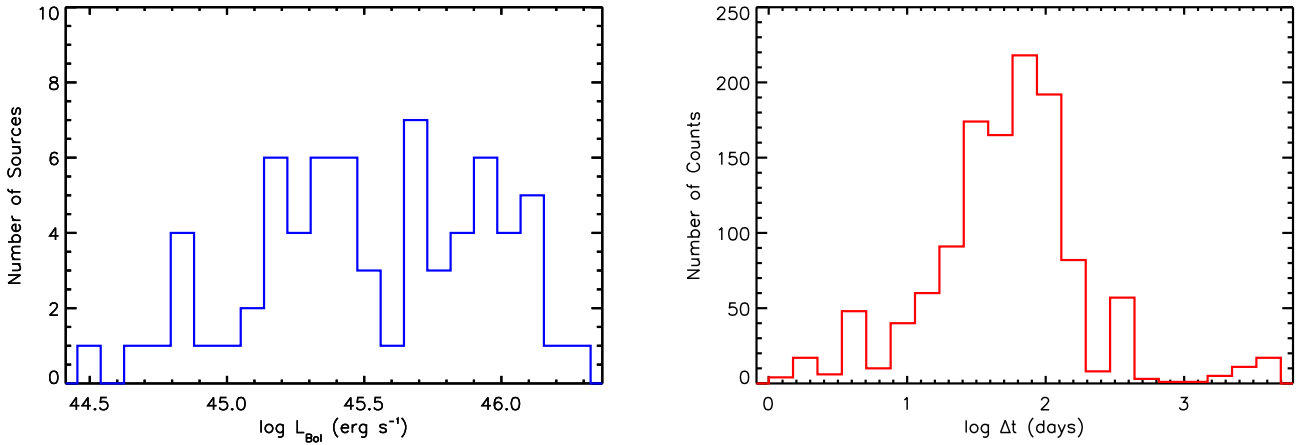}
\caption{Left: distribution of the log$L_{\rm{bol}}$ for our sample of 68 broad-line quasars. The range of  the bolometric luminosities of the quasars is 44.49 $\leq$ log$L_{\rm{bol}} <$ 46.31 and the median is log$L_{\rm{bol}} \sim$ 45.48. Right: distribution of the rest-frame $\Delta \rm{t}$ for our dataset of 1,210 points. A large fraction ($\sim$ 75\%) of the timescales are located in the region of $\Delta \rm{t} \leq100$ days. \label{f01}}
\end{figure*}

For each quasar, we estimate the bolometric luminosity using the continuum luminosity at 3000 \AA \ and $L_{\rm{bol}}=5L_{\rm{3000}}$ (e.g., \citealt{RichardsEtAl2006}). The range of the bolometric luminosities of our quasars is 44.49 erg s$^{-1} \leq \rm{log}$$L_{\rm{bol}} \leq 46.31$ erg s$^{-1}$, and the median is log$L_{\rm{bol}} \sim$ 45.48 erg s$^{-1}$. 
The distribution of the log$L_{\rm{bol}}$ of our quasars and the timescales $\Delta t$ of our dataset is shown in Figure \ref{f01}. Note that a small fraction ($\sim$ 30\%) of the SDSS-I/II quasars observed with SDSS-III is also taken into account.

\section{Spectral measurements} \label{sec:sm}

In this section, we proceed to measure the strength of $\mgii \ \lambda2798$ emission-line, the 3000 \AA \ continuum, and narrow-lines including $\nev, \oii, \neiii$ and $\oiii$. 
To derive the accurate flux of $\mgii \ \lambda2798$ and that of the 3000 \AA \ continuum emission, we used a pseudo-continuum model to fit the quasar spectra. This model consists of $\feii$ multiplets, the power-law (PL) continuum, Balmer continuum, and high order Balmer lines. Before the local fits, we corrected the Galactic extinction in the SDSS spectra using the Milky Way (MW) reddening law derived by \citet{CardelliEtAl1989} and the derived $E(B-V)$ based on \citet{SchlegelEtAl1998} dust map, and shifted the spectra to rest-frame using the cataloged redshift as the systemic redshift. Following \cite{ShenEtAl2012}, we masked out narrow absorption lines for each source to reduce the uncertainties of our continuum and emission-line fits.

Though a $\feiii$ template was derived from UV spectrum of I Zw 1 (\citealt{VestergaardEtAl2001}), we do not include it in our pseudo-continuum model because it is routinely difficult to constrain this component in spectral fits (see, e.g., \citealt{GreeneEtAl2010}). The $\feii$ template used in this work is exactly the same as the template \cite{ShenEtAl2012} used. That is, the UV $\feii$ template is a combination of templates of \citet{VestergaardEtAl2001} in 1000-2200 \AA, \citet{SalvianderEtAl2007} in 2200-3090 \AA \ and \citet{TsuzukiEtAl2006} in 3090-3500 \AA, and the optical $\feii$ template is \citet{BorosonEtAl1992} template (3686-7484 \AA). We independently fitted these two $\feii$ templates, each with three free parameters, i.e., the normalization, the velocity dispersion, and the wavelength shift of the template. For the PL continuum model, the normalization factor and the slope are used as two free parameters.

For the contribution of the Balmer continuum, we follow the formula from \citet{DietrichEtAl2003} and \citet{TsuzukiEtAl2006}, in which the Balmer continuum is expressed by
\begin{equation}
f_{BC}(\lambda)=f_{3646} B_{\lambda}(T_e)(1-e^{-\tau_{\lambda}}); \qquad \qquad \rm{if} \ \lambda \leqslant \lambda_{BE}
\end{equation}
where $f_{3646}$ is the normalization coefficient at Balmer edge 3646 \AA \ and $\tau_{\lambda}=\tau_{BE}(\lambda/\lambda_{BE})^3$ in which $\tau_{BE}$ is the optical depth at Balmer edge $\lambda_{BE}$ (3646 \AA), and $B_{\lambda}(T_e)$ is the Planck function at an electron temperature $T_e$, which is assumed to be 15,000 K (e.g., \citealt{DietrichEtAl2003}, \citealt{JiEtAl2012}). 

To improve the local fit for each narrow-line, high order Balmer lines up to $n =$ 50 are also included in our pseudo-continuum model, as \citet{JiEtAl2012} did. Utilizing the Balmer line emissivities for Case B, $T_e = 15,000$ K and $n_e = 10^8$ cm$^{-3}$ (\citealt{StoreyEtAl1995}), we constrain the relative strengths of these lines. We fix the flux ratios of the high order Balmer lines to the Balmer continuum flux at the edge (3646 \ \AA) according to the results in \citet{WillsEtAl1985}. Below we describe the detailed fitting procedures for $\mgii$ broad line, the 3000 \AA \ continuum, and several narrow lines.


\subsection{$\mgii$} \label{subsec:mgii}

For the broad $\mgii \ \lambda2798$ line, we first fitted the pseudo-continuum model consisting of the PL continuum, Balmer continuum, and the UV $\feii$ template. All these components were fitted simultaneously in the following windows: 2155-2675 \AA \ and 2925-3500 \AA, which are devoid of strong emission lines. 

We then subtracted the pseudo-continuum from the original SDSS spectra, and fitted the $\mgii$ line over the [2690,2910] \AA \ wavelength range. The broad $\mgii$ component was modelled with multiple Gaussians with up to three Gaussians (each with FWHM $\geq$ 900 km/s). As to the narrow component of $\mgii$, we used two Gaussians\footnote{Given that $\mgii \ \lambda 2798$ is a fairly widely separated doublet at 2795.530 \AA \ and 2802.704 \AA, a separation of $\sim$ 750 km s$^{-1}$ in Doppler shift, and $\mgii$ is often the narrowest of the broad emission lines, we apply the following additional constraints to the narrow component: FWHM $<$ 900 km s$^{-1}$ and flux $<$ 10\% of the total $\mgii$ flux, see, e.g., \citet{WillsEtAl1993}; \citet{McLureEtAl2004}; \citet{WangEtAl2009}. For narrow lines such as $\neiii$, $\oii$, $\neiii$, $\oiii$, the upper limit for each line width is 1200 km s$^{-1}$, see, e.g., \citet{HaoEtAl2005}; \citet{ShenEtAl2011, ShenEtAl2012}.} and checked both possibilities of doublet ratio, 2:1 for optically thin and 1:1 for optically thick, and then chose the one that has a smaller reduced $\chi^2$ value.

\subsection{$The \ 3000 \AA \ Continuum$} \label{subsec:cont}

We used the PL continuum model in the above pseudo-continuum fitting procedure to estimate the 3000 \AA \ continuum luminosity $L$ = $\lambda L_{\lambda}$ at 3000 \AA \ (\citealt{ShenEtAl2012}). We found that our local fits for pseudo-continuum and $\mgii$ line are perfect and the reduced $\chi^2$ values from the spectral fits are close to 1, with the median value of 1.15 and 0.98.

\subsection{$\nev$\label{subsec:ne5}}

For $\nev \ \lambda3346,3426$ doublet, the pseudo-continuum model fitting wavelength windows are [2480,2675] \AA, [2925,3020] \AA, [3225,3300] \AA \ and [3450,3550] \AA. These wavelength coverages are not contaminated with strong emission lines. After subtracting the pseudo-continuum from the spectrum, we fit the wavelength range [3329,3446] \AA \ for $\nev$. Considering the spectroscopic S/N of $\nev$, we used two Gaussians for the $\nev \ \lambda3346$ and $\nev \ \lambda3426$ lines, and we tied their flux ratio to be $f_{3426}/f_{3346} = 3$ during the fit.

Since $\nev$ doublet are intrinsically weak lines, we did not adopt such restrict condition as S/N $\geq$ 10 in our spectral fitting procedure. Instead, we discard the rescaling factor differing significantly from the ones obtained from other narrow-lines in the spectra. Nevertheless, the doublet could provide reference to the flux-calibration correction factors that are calculated from other narrow-lines. (see Section \ref{subsec:vc} for more details).
 
\subsection{$\oii$ \ $\& \ \neiii$\label{subsec:o2ne3}}

For $\oii \ \lambda3728$ and $\neiii \ \lambda3869,3968$ lines, as there are no strong nearby broad lines, we simultaneously fitted the PL continuum, $\feii$ emission (the optical $\feii$ template), high order Balmer lines, $\hei \ \lambda$3889, $\hf \ \lambda$3890, $\he \ \lambda$3971, $\oii$ and $\neiii$ over the wavelength range 3670-4020 \AA. The narrow components of these emission lines were each fit with a single Gaussian. On the other hand, the broad components of $\hf \ \lambda3890$ and $\he \ \lambda$3971 were each fit with multiple Gaussians up to three Gaussians. 
We tied the flux ratio of the $\neiii \ \lambda \lambda 3869,3968$ doublet to be $f_{3869}/f_{3968}$ = 3 during the fitting.

Note that the $\oii$ narrow-line is comprised of the blended $\oii \ \lambda \lambda$3726,3729 doublet, which is rarely resolved in the SDSS spectra due to the inadequate spectral quality (for instance, do not have adequate S/N or low-resolution spectroscopy to unambiguously locate the doublet). We tried to fit the wavelength range [3680,3780] \AA using up to two Gaussians for the subtracted spectra (i.e., leaving $\oii$ and $\neiii$ emission-lines) when the spectroscopic S/N ($\oii$) $\geq$ 10. It is shown that the final $\chi^2$/dof value from our fitting procedure for each spectrum is usually close to 1. 

\subsection{$\oiii$\label{subsec:o3}}

For $\oiii \ \lambda4960,5008$ doublet, we first used the optical $\feii$ template and the PL continuum fit for each object. The PL continuum$+$iron fitting windows are [4435,4700] \AA \ and [5080,5535] \AA. We then subtracted the pseudo-continuum from the spectrum (leaving the emission-line spectrum). Considering the strong contamination of $\hb \ \lambda4861$ broad-line on $\oiii$ lines, we simultaneously fitted $\hb$ line and the pseudo-continuum model. We fitted the wavelength range [4700,5100] \AA \ and used one Gaussian with FWHM $<$ 1200 km/s for the narrow $\hb$ component and up to three Gaussians (each with FWHM $\geq$ 1200 km/s) for the broad $\hb$ component. 

Considering the modest quality of the SDSS spectra, we fitted each of the two $\oiii$ lines with a single Gaussian. We tied the flux ratio of the $\oiii$ doublet to be $f_{5008}/f_{4960}$ = 3 to reduce possible ambiguities. It is true that asymmetric blue wings (see, e.g., \citealt{HeckmanEtAl1981}; \citealt{KomossaEtAl2008}) and dramatic double-peaked profiles (e.g., \citealt{LiuEtAl2009}; \citealt{WangEtAl2009}) often appeared in the $\oiii$ lines. However, these features can only be well constrained in high S/N ($\geq$ 10) spectra. For these spectra, we used up to two Gaussians for each of the narrow $\oiii$ lines.

\subsection{Measurement Uncertainties\label{subsec:mu}}

It is of immense significance to determine the uncertainties in the continuum and emission-line flux measurements. We followed \citet{ShenEtAl2012} and adopted the following Monte-Carlo method:

\begin{enumerate}
\item Using the given flux density errors, we perturb the original spectra randomly to generate mock spectra.
\item The mock spectra are each fitted with the same fitting routine to get the corresponding line fluxes.
\item We repeat 1 and 2 for 50 times and obtain a distribution of our spectral measurements.
\item The semi-amplitude of the range enclosing the 16th and 84th percentiles of the distribution is adopted as an estimation of our measurement uncertainties.
\end{enumerate}

Figures 2 and 3 show two examples of our fits to the spectra in presence of different levels of UV/optical $\feii$ emission.
\begin{figure*}[t]
\begin{tabular}{llll}
\begin{minipage}[t]{3.4in}
\includegraphics[width=3.4in]{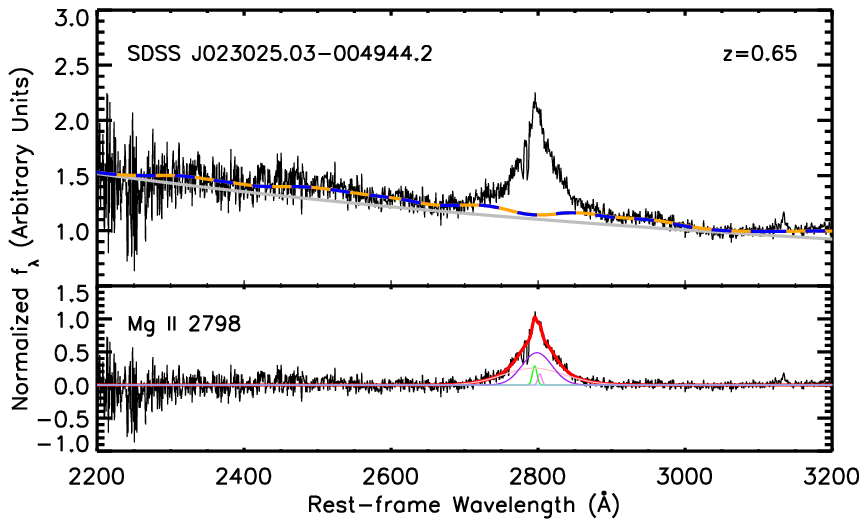}
\end{minipage}
\begin{minipage}[t]{3.4in}
\includegraphics[width=3.4in]{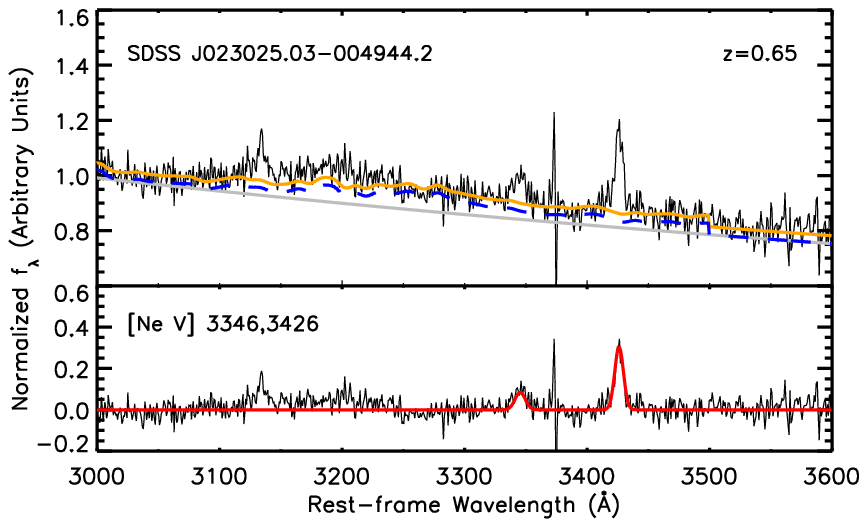}
\end{minipage}
\\
\begin{minipage}[t]{3.4in}
\includegraphics[width=3.4in]{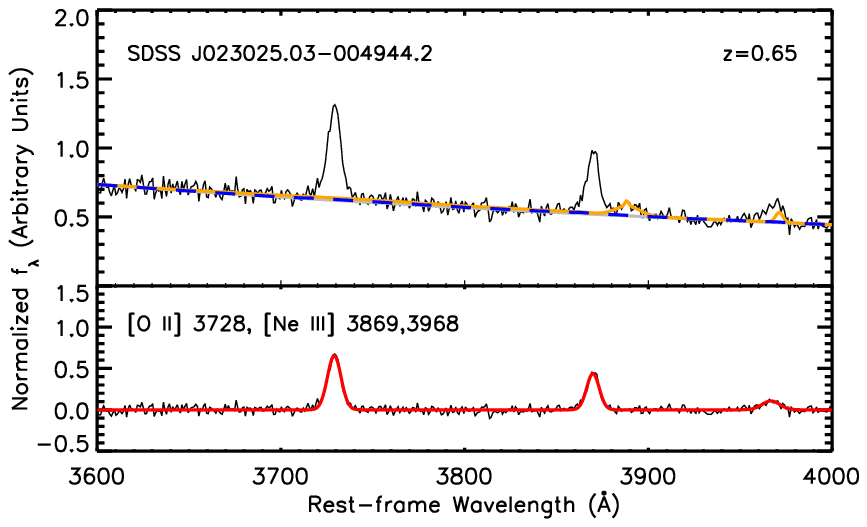}
\end{minipage}
\begin{minipage}[t]{3.4in}
\includegraphics[width=3.4in]{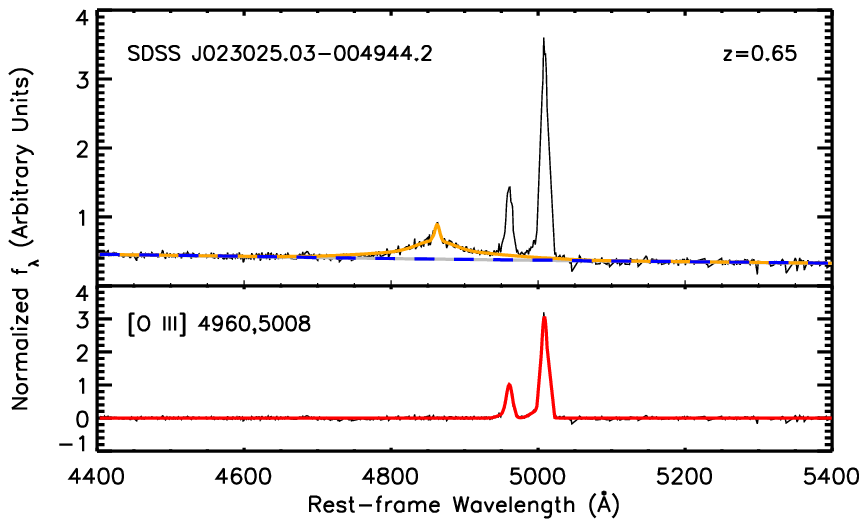}
\end{minipage}
\end{tabular}
\caption{Example of our model fits to different emission-lines in the spectra of a broad-line quasar (J023025.03-004944.2). The original spectrum (black), the power-law continuum (gray), the continuum+$\feii$ template fit (blue dashed), and the combined pseudo-continuum model (orange) to be subtracted off are presented in the upper panel of each plot. The corresponding bottom panels show the emission-line fits to $\mgii$ through $\oiii$ doublet, where black lines are the residuals and red lines are the combined model line profiles. For $\mgii$, we also show the model narrow-line emission in green and violet and the model broad-line emission in purple, pink and skyblue. \label{f02}}
\end{figure*}
\begin{figure*}[t]
\begin{tabular}{llll}
\begin{minipage}[t]{3.4in}
\includegraphics[width=3.4in]{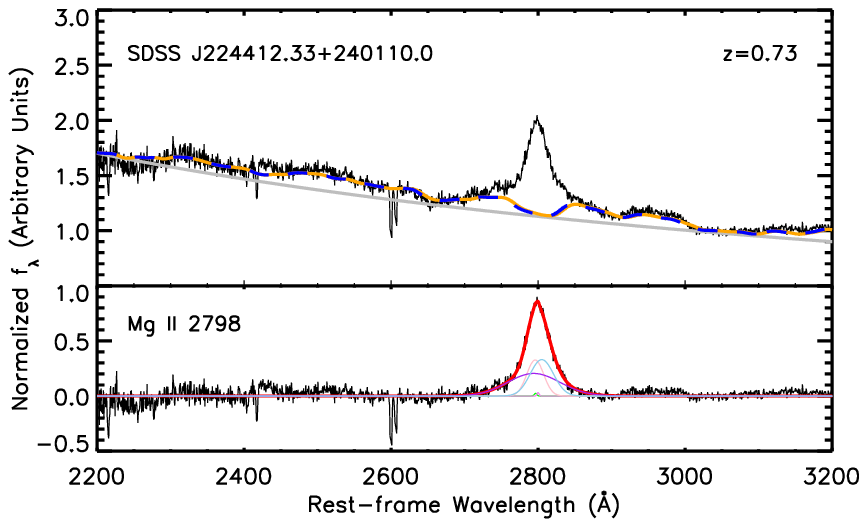}
\end{minipage}
\begin{minipage}[t]{3.4in}
\includegraphics[width=3.4in]{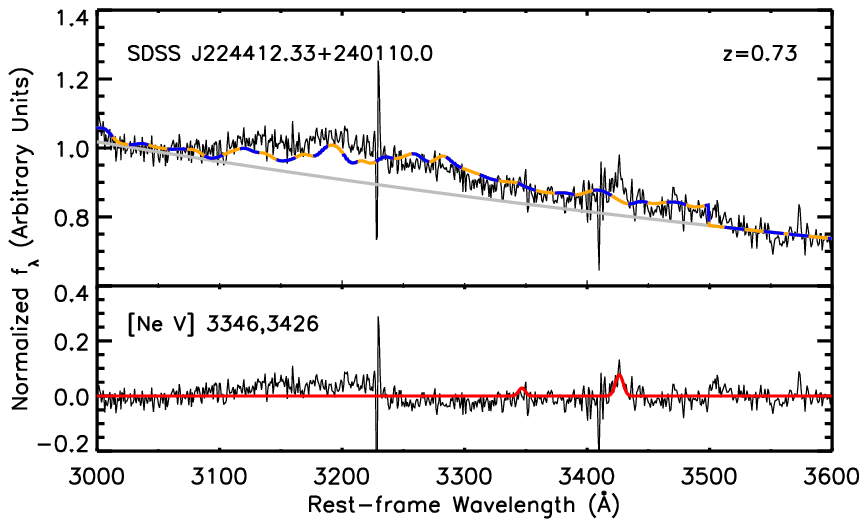}
\end{minipage}
\\
\begin{minipage}[t]{3.4in}
\includegraphics[width=3.4in]{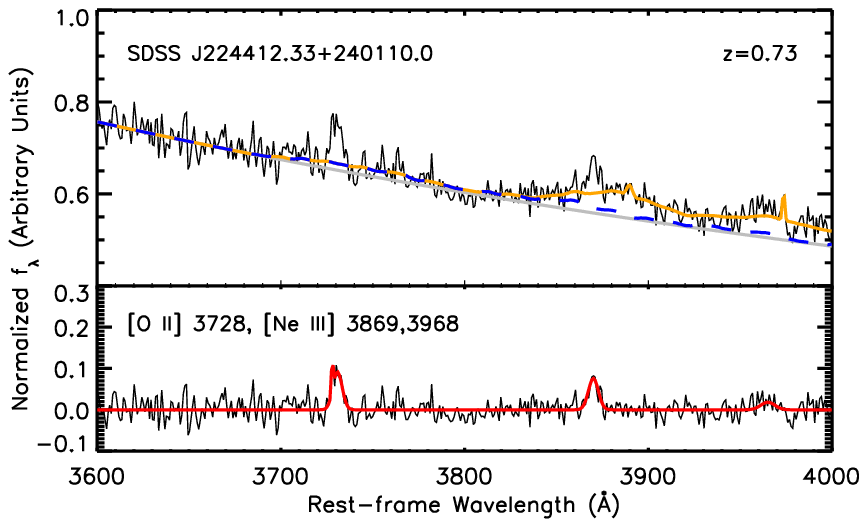}
\end{minipage}
\begin{minipage}[t]{3.4in}
\includegraphics[width=3.4in]{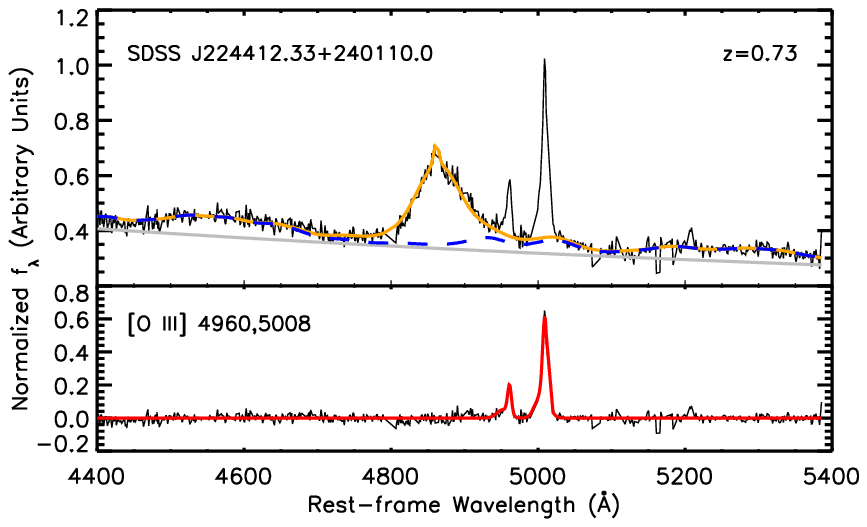}
\end{minipage}
\end{tabular}
\caption{Another example of our model fits to different emission-lines in the spectra of a broad-line quasar (J224412.33+240110.0) that involves stronger UV/optical $\feii$ emission. The original spectrum (black), the power-law continuum (gray), the continuum+$\feii$ template fit (blue dashed), and the combined pseudo-continuum model (orange) to be subtracted off are presented in the upper panel of each plot. The corresponding bottom panels show the emission-line fits to $\mgii$ through $\oiii$ doublet, where black lines are the residuals and red lines are the combined model line profiles. For $\mgii$, we also show the model narrow-line emission in green and violet and the model broad-line emission in purple, pink and skyblue. \label{f002}}
\end{figure*} 
We visually inspected our fits and rejected BALQSOs.


\section{Results} \label{sec:results}


\subsection{Variations Calculation}\label{subsec:vc} 
Now we start to determinate the flux variations of broad $\mgii$ and of continuum. During the calculation, we adopt narrow-line flux-recalibration to the line flux variations. Following the conversion between fluxes and magnitudes and applying the narrow-line flux rescaling factor to the flux of broad $\mgii$ and continuum for each source, we define the basic variation as:

\begin{equation}
\Delta m = -2.5 \mathrm{log} (f_2/f_1)+\Delta m_{\overline{r}}
\end{equation}
where $f_{1}$,$f_{2}$ denote the observed fluxes of broad $\mgii$ or continuum at two epochs, $\Delta m_{\overline{r}}$ represents the variation of our final rescaling factor cross the epochs, and $r$ is the rescaling factor from each narrow-line during two observations. 

We calculate $\Delta m$ for the flux pairs of broad $\mgii$ and the 3000 \AA \ continuum separated by $\Delta t$ for each source. Assuming that one quasar was observed with $n$ spectroscopic epochs, the largest number of our data points (i.e., $\Delta m$ pairs) for this object would be $C_n^2$, if we take no account of the spectral quality. 
For each $\Delta m$ pair, we chose the narrow-line rescaling factors in the range of 0.6$-$1.4. These data pairs occupy the vast majority of our dataset. 

Given that most data points have different rescaling factors for different narrow-line, making the selection of the rescaling factors becomes one of keys in our variation calculation. To obtain a fairly reliable rescaling factor for each data point, we calculate error-weighted average value of all narrow-line rescaling factors (i.e., each rescaling factor is weighted by its uncertainty). In addition, we also require that the final rescaling factors is still in the range of 0.6$-$1.4, and the corresponding error is limited to 0.10. As a consequence, for the data pair with at least two S/N $\geq$ 5 narrow-lines, the median of the standard deviation from the mean (i.e., final correction factor) for these data pairs is $\sim$ 0.05, reflecting that the different narrow-line rescaling factors for each data pair are pretty tightly bunched together; for the data pair with only one S/N $\geq$ 5 narrow-line, our limited rescaling factor error requires that the S/N of this narrow-line should be $\geq$ 14, which is good enough to do flux-recalibration. The total number of our data points for the following variations correlation analysis is 1210 (see Section \ref{subsec:vcc} and \ref{subsec:sbc}).

\begin{figure}
\includegraphics[angle=0,scale=1.0]{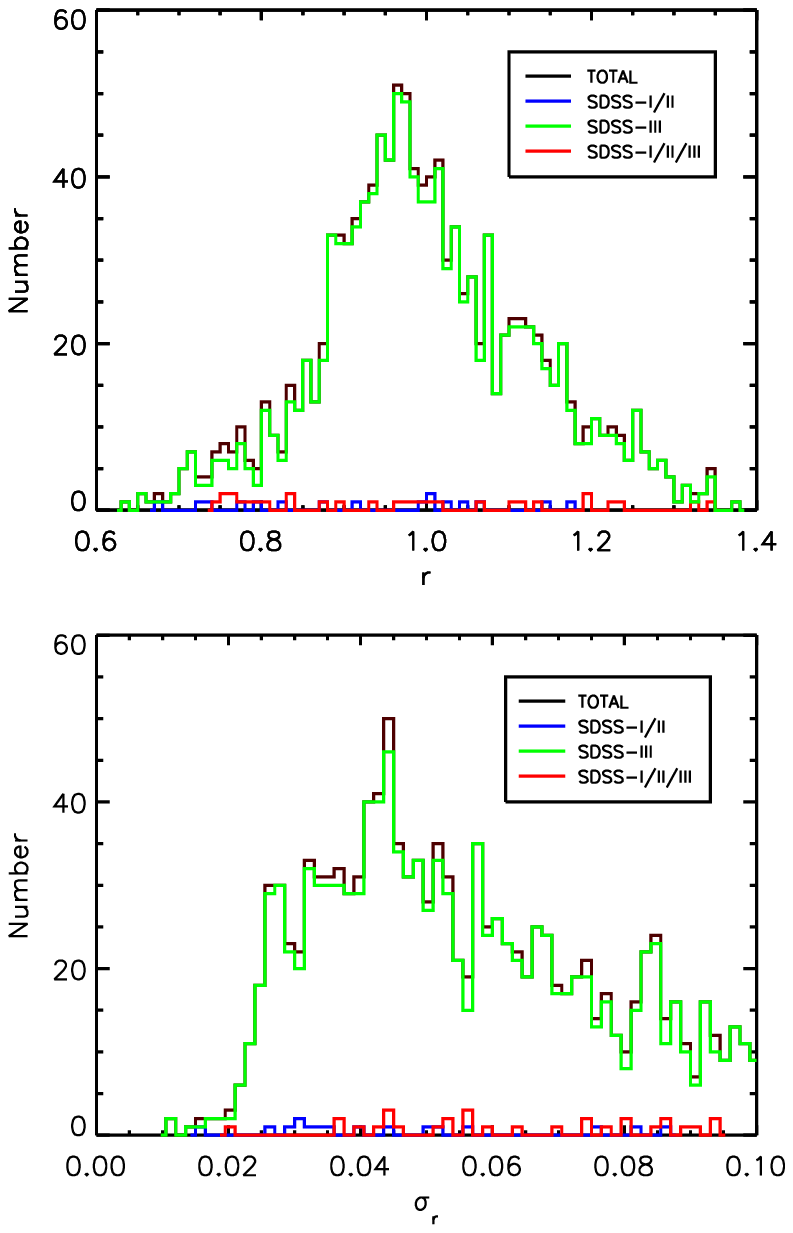}
\caption{Distribution (top panel) and error distribution (bottom panel) of the final narrow-line flux rescaling factor for the epochs in SDSS-I/II (blue), SDSS-III (green), and SDSS-I/II/III (red) in variation calculation for all the quasars in our sample. The median are 0.98 and 0.05, respectively. \label{f03}}
\end{figure}

Figure \ref{f03} shows the distribution and error distribution of our final rescaling factor in our variation calculation for all the data pairs.
We notice that \citet{ShenEtAl2016} obtained a precision ($\sim$ 5\%) of the spectroscopy achieved for SDSS-RM in reference to the median asbolute deviation (MAD) of narrow-line flux variations; while here we try not to make deductions from this parameter, as we did a totally different rescaling factor measurement. More important, our sample includes the quasars observed with both SDSS-I/II and SDSS-III, where studies have confirmed a deficit in flux in BOSS of roughly 20\% relative to SDSS-I/II at long wavelengths (see, e.g., \citealt{HarrisEtAl2016}).

\subsection{Variations Correlation Coefficient}\label{subsec:vcc}

\begin{figure*}
\includegraphics[angle=0,scale=0.985]{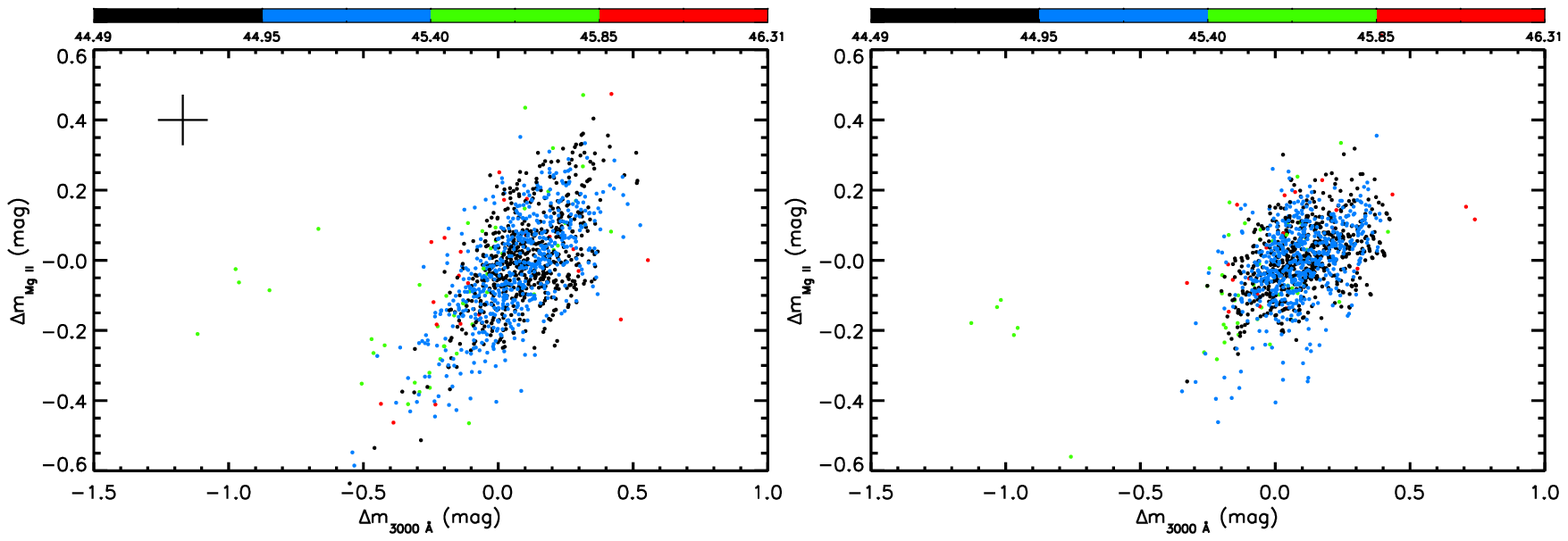}
\caption{Left: the narrow-line flux-recalibrated variations of $\mgii\ \lambda2798$ against those of the 3000 \AA \ continuum on timescales of $\Delta t \geqslant$ 1 day for our quasar sample. The typical errors are shown in the upper left corner. Right: same as the left panel, but the variations of $\mgii$ and those of the 3000 \AA \ continuum are not corrected using the constancy of the narrow-lines. We do not show the typical error in the right panel, as our sample includes a set of quasars with observations in both SDSS-I/II and SDSS-III surveys.\label{f04}}
\end{figure*}

First we compare the narrow-line flux-recalibrated variations in broad $\mgii$ with those in the 3000 \AA \ continuum for all the quasars in our sample in the left panel of Figure \ref{f04}. 
For comparison, we also plot uncorrected line and continuum variations on the same timescales in the right panel. The correlation between the narrow-line flux-recalibrated variations in broad $\mgii$ and the 3000 \AA \ continuum is tested using the Spearman rank correlation test. The null hypothesis is that there is no correlation between the input datasets. The correlation coefficient of this test is $\rho$ = 0.593 (0.456 for the dataset without narrow-line flux-recalibration). If we focus on the data on timescales of $\Delta t \leqslant$ 100 days, the result is 0.644 (0.453 for the other dataset), which is consistent with the result in \citet{SunEtAl2015}. 

\begin{figure*}
\includegraphics[angle=0,scale=0.97]{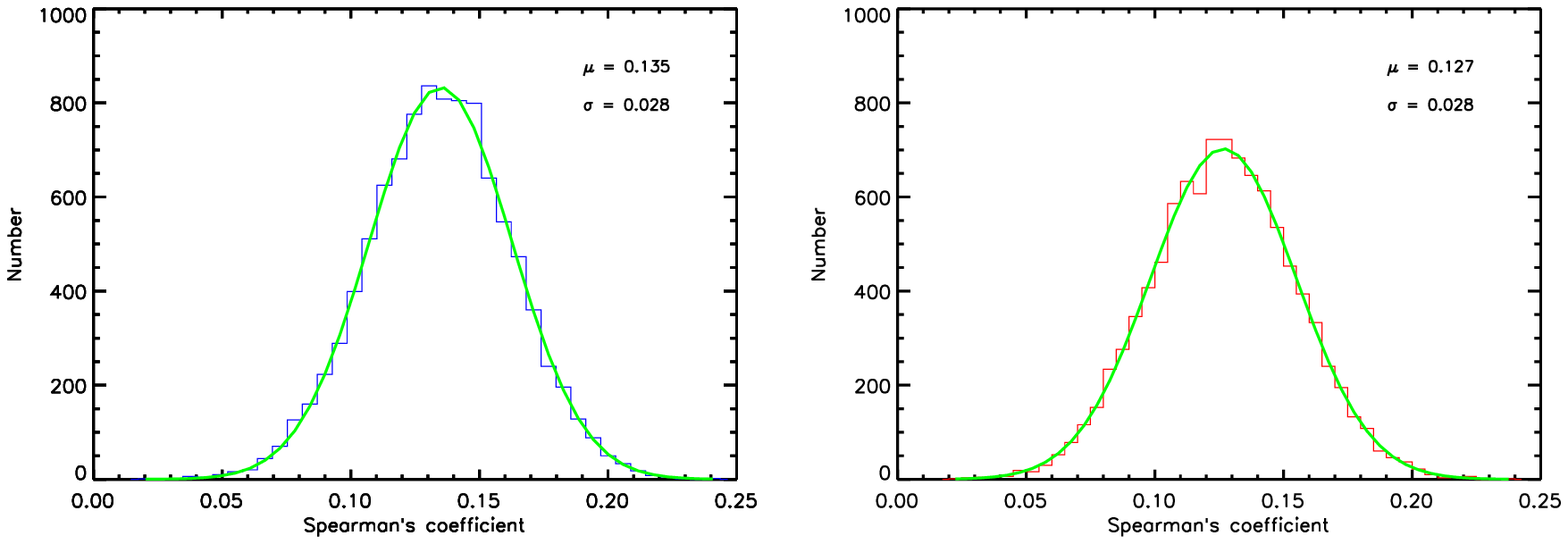}
\caption{Distribution of the Spearman's rank correlation coefficient $\rho$ of 10,000 random samples due to the correlated error of narrow-line rescaling factor. Left: the random samples are deduced from flux-recalibrated dataset. Right: the random samples are deduced from partly flux-recalibrated dataset according to one certain criterion (see the text in Section \ref{subsec:vcc}).\label{f06}} 
\end{figure*}

For the $p$ value (i.e., the probability of being incorrect in rejecting the null hypothesis), 
we adopt a bootstrap resampling method. That is, we randomly (with replacement) select $\Delta m$ pairs from the observed sample. We then perturb the data pairs of the random sample by their uncertainties.
We calculate spearman's $\rho$ for the random sample and repeat the same process for many times (e.g., 10,000) to get an estimation of the significance level. The distribution of the Spearman's $\rho$ obtained from our 10,000 random samples is shown in the left panel of Figure \ref{f06}. 
We reason that the $p$ value is far less than 0.01$\%$, indicating a significant positive correlation between the variations of broad $\mgii$ and the 3000 \AA \ continuum.

In addition, 
we also attempt to recalibrate only the variations that the narrow-line flux-recalibration errors are small enough in comparison with the rescaling factor (e.g., $|\ln r| > $2 $\Delta r$) to generate a partly recalibrated dataset. Adopting the bootstrap resampling method to the two datasets on timescales of $\Delta t \leq$ 3 days, we find that our flux-recalibration process hardly biases our variation correlation analysis by comparing both significance levels. To make further validation, we also plot the distribution of the Spearman's $\rho$ for the random samples deduced from partly flux-recalibration on the timescales of $\Delta t \geq$ 1 day in the right panel of Figure \ref{f06}. The comparison of both significance levels infers that our flux-recalibration for all variations does not introduce a much larger uncertainty for the points with relatively large rescaling factor errors in our correlation analysis.

\subsection{The Slope Between Variations of Two Components}\label{subsec:sbc}

Given the correlated measurement uncertainties of the broad $\mgii\ \lambda2798$ emission-line and the 3000 \AA \ continuum variations, we perform a modified weighted least squares regression method to calculate the slope between narrow-line flux-recalibrated variations of both components. The covariance of the $\mgii$ line and the continuum variations, generated by the bootstrap resampling method, is also taken into consideration in our modified method. Assuming a linear model, we can get the following relation.

\begin{equation}
\begin{aligned}
\Delta m_{\rm{line0}}+\Delta m_{\overline{r}} = \Delta m_{\rm{line}} = a+b*\Delta m_{\rm{cont}} \\
 =  a+b*(\Delta m_{\rm{cont0}}+\Delta m_{\overline{r}})
\end{aligned}
\end{equation}
where in this case $\Delta m_{\rm{line}}$, $\Delta m_{\rm{cont}}$ represent narrow-line flux-recalibrated variations in broad $\mgii$ and in the continuum respectively. $\Delta m_{\rm{line0}}$, $\Delta m_{\rm{cont0}}$ are the corresponding original variations without narrow-line flux-recalibration, respectively. Clearly, both the $\Delta m_{\rm{line}}$ and the $\Delta m_{\rm{cont}}$ are correlated with $\Delta m_{\overline{r}}$, which is induced by the flux calibration. The total measurement uncertainty of our linear weighted least squares fit model is as follows.

\begin{equation}
\begin{aligned}
\sigma_{tol}=\sqrt{\sigma_{\Delta m_{\rm{line}}}^2+b^2*\sigma_{\Delta m_{\rm{cont}}}^2-2*b*\sigma_{\Delta m_{\rm{line}} \Delta m_{\rm{cont}}}^2}
\end{aligned}
\end{equation}
where $\sigma_{\Delta m_{\rm{line}} \Delta m_{\rm{cont}}}^2$ is the covariance of the variations of emission line and the continuum. In the calculation of $\sigma_{\Delta m_{\rm{line}} \Delta m_{\rm{cont}}}^2$, we assign normal distributed errors to $\Delta m_{\rm{line0}}$, $\Delta m_{\rm{cont0}}$, and $\Delta m_{\overline{r}}$ for each data point, as these parameters are independent of each other. Then we get an estimation of the covariance after repeating the same process for many times (e.g., 10,000).

\begin{figure}
\includegraphics[angle=0,scale=1.0]{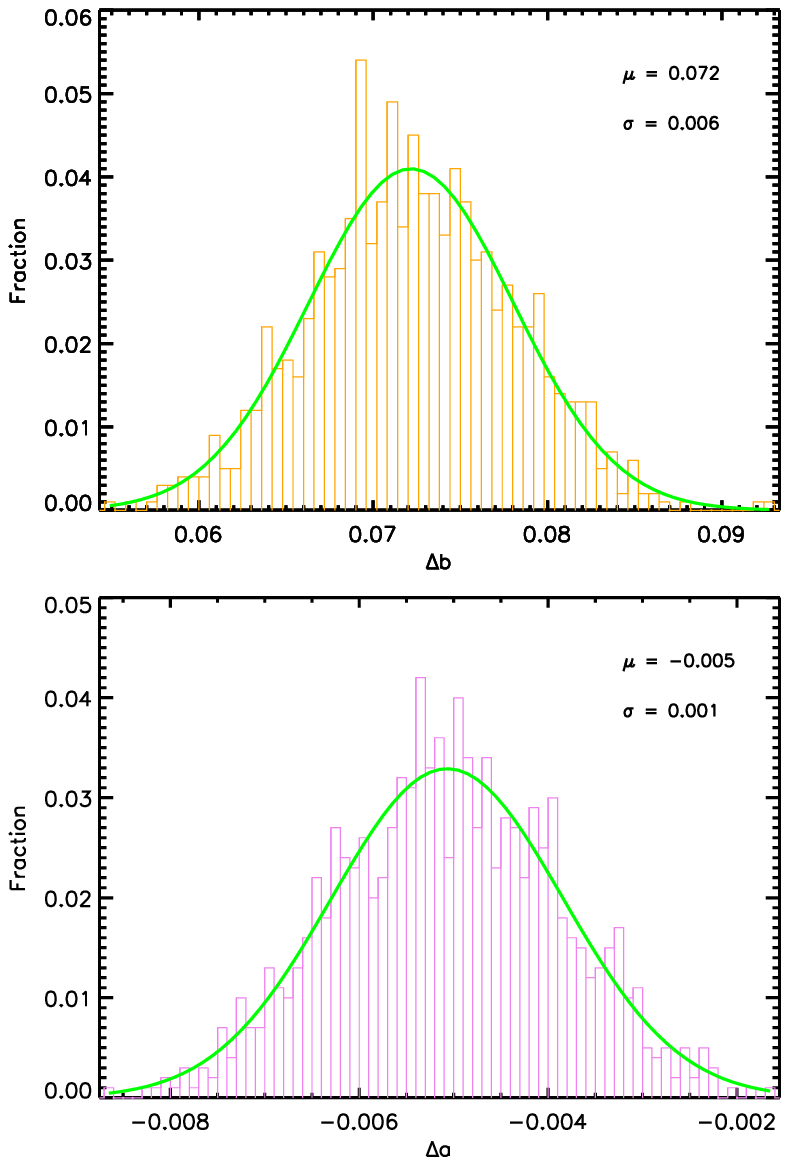}
\caption{Example distribution of slope bias (top panel) and intercept bias (bottom panel) for 100,000 random samples due to narrow-line rescaling factor error in the slope (and intercept) estimation for our quasars spanning a bolometric luminosity range of 44.95 erg s$^{-1} <$ log$L_{\rm{bol}}$ $<$ 45.40 erg s$^{-1}$.\label{f07}}
\end{figure}

Note that the variations of emission line and of the continuum are dependent on quasar luminosities, and hence we will discuss the slopes for different luminosity bins. Our sample is divided into 4 sub-samples by log$L_{\rm{bol}}$: 44.49 erg s$^{-1} \leq$ log$L_{\rm{bol}} <$ 44.95 erg s$^{-1}$; 44.95 erg s$^{-1}$ $\leq$ log$L_{\rm{bol}} <$ 45.40 erg s$^{-1}$; 45.40 erg s$^{-1}$ $\leq$ log$L_{\rm{bol}} <$ 45.85 erg s$^{-1}$; 45.85 erg s$^{-1} \leq$ log$L_{\rm{bol}} <$ 46.31 erg s$^{-1}$. Adopting our modified weighted least squares method (see equation (4)), we calculate the slope between the variations of both components for the quasars in each luminosity bin. The intrinsic bias introduced by narrow-line flux-recalibration process itself is estimated by the following method. 1) we constrained the intrinsic variability of the 3000 \AA \ continuum; 2) we assumed that variations (X) follow a Guassian distribution, whose RMS equals to the intrinsic variability of the 3000 \AA \ continuum; 3) we generated mock samples from such a distribution; 4) we calculated mock line variations by adopting Y = a + b * X; 5) we fitted the slope between these two mock variations; 6) we repeated this process for 100,000 times and obtained the distribution of the fitted slope. The difference between the mean of the slope measurement and the value of the input parameter b is then defined as the intrinsic slope bias. In Figure \ref{f07}, 
we show an example of the intrinsic slope bias ($\sim$ 0.072) due to our correction to the variations of both components for the sources covering a bolometric luminosity range of 44.95 erg s$^{-1}$ $\leq$ log$L_{\rm{bol}} <$ 45.40 erg s$^{-1}$. 

\begin{figure*}[t]
\begin{tabular}{llll}
\begin{minipage}[t]{3.43in}
\includegraphics[width=3.43in]{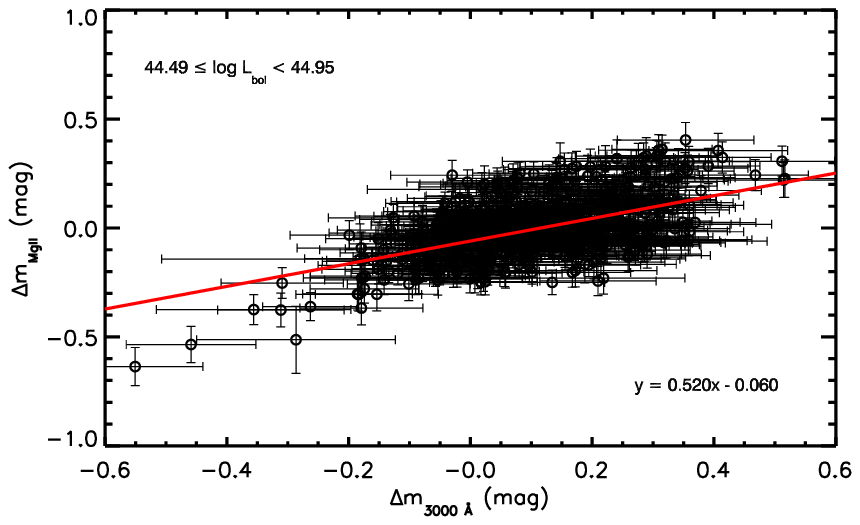}
\end{minipage}
\begin{minipage}[t]{3.43in}
\includegraphics[width=3.43in]{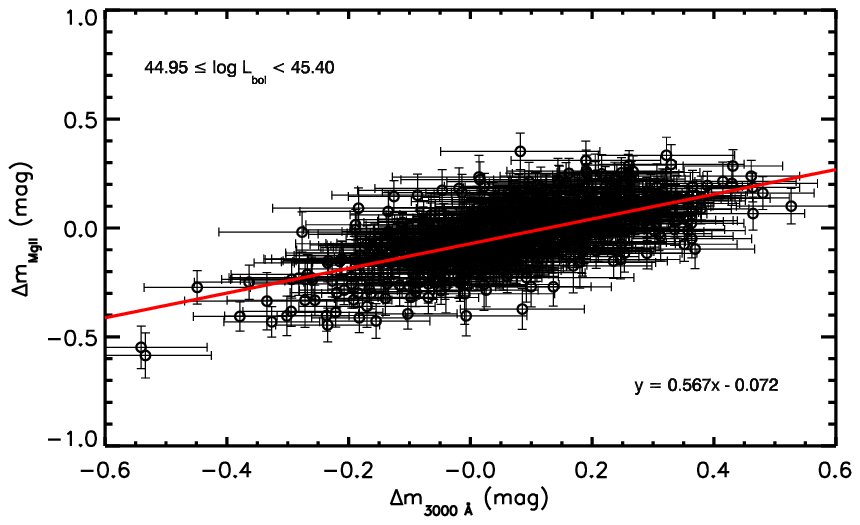}
\end{minipage}
\\
\begin{minipage}[t]{3.43in}
\includegraphics[width=3.43in]{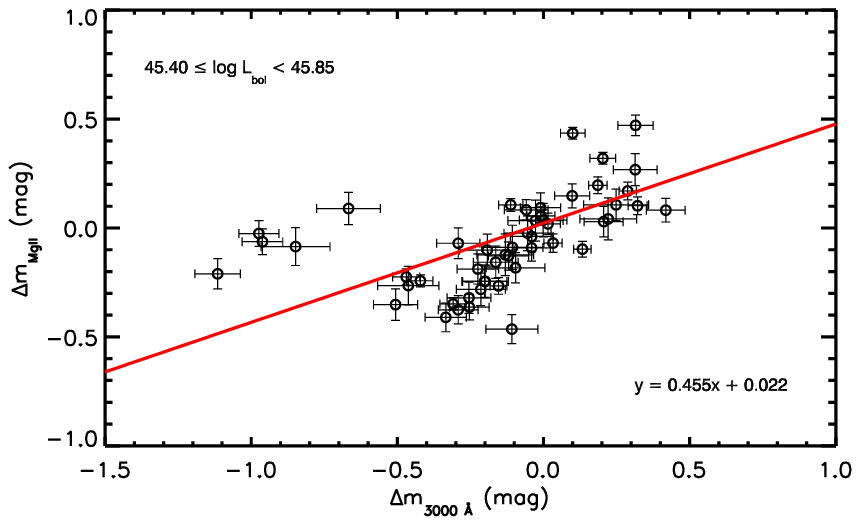}
\end{minipage}
\begin{minipage}[t]{3.43in}
\includegraphics[width=3.43in]{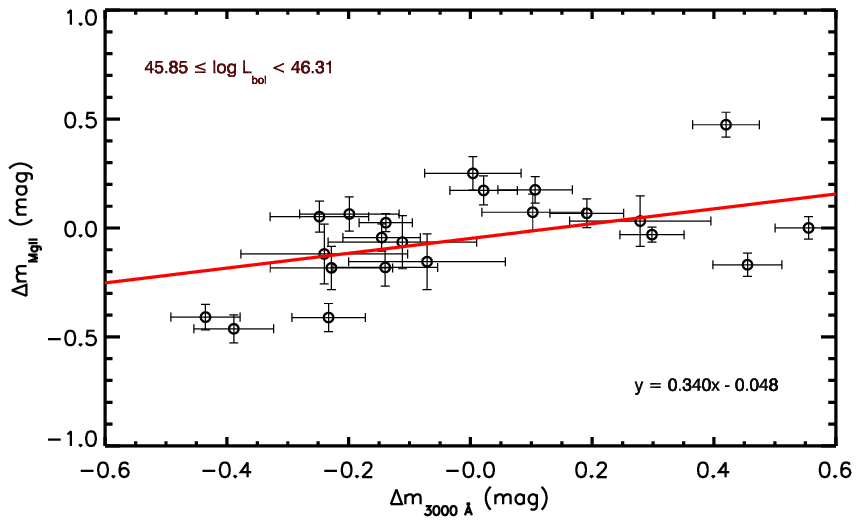}
\end{minipage}
\end{tabular}
\caption{Comparisons between the narrow-line flux-recalibrated variations of $\mgii$ and of the 3000 \AA \ continuum for the quasars in different bolometric luminosity bins. The solid red curve in each plot is a modified linear weighted least squares fit to our data. \label{f08}}
\end{figure*}

After eliminating the intrinsic biases introduced by the rescaling process, we obtain the corrected slopes between the narrow-line flux-recalibrated variations of broad $\mgii$ and of the 3000 \AA \ continuum for our quasars in different luminosity bins. Table \ref{tb2} presents our parameter measurements of the correlation between variations of $\mgii$ line and continuum. It is revealed that the slope decreases as the quasar luminosity increases. In Figure \ref{f08}, 
we compare the emission line variations with the continuum variations for our 4 sub-samples. We also estimate virial SMBH masses using the 3000 \AA \ continuum luminosity and archive $\mgii$ FWHM data from SDSS-DR7Q and BOSS-DR12Q (e.g., \citealt{ShenEtAl2011}; \citealt{SunEtAl2015}). The median SMBH mass for each luminosity bin is $\sim$ constant (approximately $\sim 10^8$ $M_{\odot}$), meaning that the slope and Eddington ratio are also anti-correlated.
Additionally, we obtained an distribution of $\Delta m_{_{\rm{Mg II}}}/ \Delta m_{\rm{3000 \AA}}$ (equivalent to the responsivity $\alpha$) for our 1210 data pairs of spectroscopic observations of 68 SDSS quasars. The median $\alpha$ = 0.428 is roughly consistent with our average corrected slope (0.464 $\pm$ 0.013). 


\section{Discussion} \label{sec:dis}
There have been some studies discussing the correlation between emission-line and continuum variations (e.g., \citealt{Baldwin1977};  \citealt{KinneyEtAl1990}; \citealt{PoggeEtAl1992}; \citealt{GoadEtAl1993}; \citealt{OBrienEtAl1995}; \citealt{GilbertEtAl2003}; \citealt{GoadEtAl2004}; \citealt{KongEtAl2006}). However, most of these correlation studies either focused on other emission-lines (e.g., $\lya \ \lambda1216$, $\civ \ \lambda1549$, $\hb \ \lambda4861$) or did not correct the ensemble flux variations using narrow-line fluxes. Thus we attempt to statistically investigate the relationship between the flux variations of broad $\mgii\ \lambda2798$ emission-line and of the nearby 3000 \AA \ continuum using narrow-line calibrated SDSS spectra.


Our results generally agree with earlier observational studies that the variations in the $\mgii \ \lambda2798$ emission-line is well correlated with those in the 3000 \AA \ continuum (e.g.,\citealt{WilhiteEtAl2005}; \citealt{Woo2008}; \citealt{BenitezEtAl2009}; \citealt{CackettEtAl2015}; \citealt{SunEtAl2015}). We analysed the data from the SDSS-RM project in \citet{SunEtAl2015} using the same method (i.e., modified weighted least squares regression) and found that the slope between the variations of broad $\mgii$ and of the 3000 \AA \ continuum for this dataset is 0.620 $\pm$ 0.028. Though different quasar samples are used in the two studies, both
\begin{deluxetable*}{ccccccccc}
\tabletypesize{\footnotesize}
\tablecaption{\label{tb2} Relation Parameter Measurements}
\tablewidth{0pt}
\tablehead{
\colhead{ID} & \colhead{Luminosity Bin} & \colhead{Spearman's $\rho$} &  \colhead{Fitted Slope} & \colhead{$\Delta t$ Range} & \colhead{Median $M_{\bullet}$} &
\colhead{$N_{\rm{obj}}$} & \colhead{Slope Bias} & \colhead{Corrected Slope}
\\
\colhead{(1)} & \colhead{(2)} & \colhead{(3)} & \colhead{(4)} & \colhead{(5)} &
\colhead{(6)} & \colhead{(7)} & \colhead{(8)} & \colhead{(9)} 
}
\startdata
1 & 44.49 $\leq$ log$L_{\rm{bol}} <$ 44.95 &  0.485 &  0.556 $\pm$ 0.026 & $\leq$727 & 7.6x$10^7$ & 8 & 0.036 & 0.520 $\pm$ 0.026 \\
2 & 44.95 $\leq$ log$L_{\rm{bol}} <$ 45.40 & 0.658 &  0.639 $\pm$ 0.031 & $\leq$4035 & 1.3x$10^8$ & 21 & 0.072 & 0.567 $\pm$ 0.031 \\
3 & 45.40 $\leq$ log$L_{\rm{bol}} <$ 45.85 & 0.697 &  0.467 $\pm$ 0.025 & $\leq$4781 & 2.8x$10^8$ & 21 & 0.012 & 0.455 $\pm$ 0.025 \\
4 & 45.85 $\leq$ log$L_{\rm{bol}} <$ 46.31 & 0.526 & 0.373 $\pm$ 0.050 & $\leq$3880 & 3.7x$10^8$ & 18 & 0.033 & 0.340 $\pm$ 0.050 \\
\dag & 44.49 $\leq$ log$L_{\rm{bol}} <$ 46.31 & 0.593 & 0.513 $\pm$ 0.013 & $\leq$4781 & 2.3x$10^8$ & 68 & 0.049 & 0.464 $\pm$ 0.013 \\
\enddata
\tablecomments{Column 1: identification number assigned in this paper. Column 2: divided quasar luminosity bins. Colomn 3-9: the Spearman's coefficient, fitted slope, range of timescale (in units of days), median quasar virial SMBH mass (in units of $M_{\odot}$), total number of sources, the intrinsic fitted slope bias due to the narrow-line rescaling factor, and the corrected slope that is equivalent to the responsivity $\alpha$ of $\mgii$ for the quasars in each luminosity bin.}
\end{deluxetable*}
 results indicate a very small value in the responsivity of $\mgii$.
This suggests that the line not be expected to respond strongly to changes in continuum flux, and that it is might not easy to detect a plausible lag between the 3000 \AA \ continuum and the $\mgii$ variations (see, e.g., \citealt{CackettEtAl2015}).


On the other hand, using a sample of 101 quasars and 88 Seyferts with multiple International Ultraviolet Explorer (IUE) observations, \citet{KinneyEtAl1990} confirmed the existence of a correlation between the continuum and the $\civ \ \lambda1549$ equivalent width with a slope of $\beta \sim$ -0.17 $\pm$ 0.04. They further concluded a similar relation for $\lya \ \lambda1216$ emission-line with a slope of $\beta \sim$ -0.12 $\pm$ 0.05. Given that $\alpha$ = $\beta$ + 1, the responsivity $\alpha$ of $\civ$ is $\sim$ 0.83 $\pm$ 0.04 and that of $\lya$ is $\sim$ 0.88 $\pm$ 0.05. This indicates that, unlike $\mgii$, $\civ$ and $\lya$ vary greatly in response to the variability of the continuum emission. Subsequently, \citet{ GilbertEtAl2003} and \citet{GoadEtAl2004} studied the $\hb$ variability for NGC 5548 using the IUE archived UV spectra and/or ground-based optical monitoring data of 13-year-observations. They found that the responsivity $\alpha$ of the broad $\hb \ \lambda4861$ varies from 0.4 (bright states) to 1 (dim states), anti-correlated with the flux of the incident continuum. This is consistent with our result of a negative relationship or inverse relationship between the emission-line responsivity and Eddington ratio.

In addition, our results are roughly in good agreement with theoretical studies on the intrinsic variability of $\mgii$. \citet{GoadEtAl1993} computed the response function for 
$\mgii \ \lambda2798$ covering various ionization states originated from the spherical BLR. Assuming a panchromatically changing flux in the incident continuum, they found that the responsivity of the $\mgii$ emission-line varies throughout the BLR, but is generally small ($\alpha <$ 0.5).
Given that the shorter wavelength UV continuum usually has larger amplitude variability relative to the continuum at longer wavelength, our estimation of $\alpha$ might be slightly higher compared to (but roughly consistent with) the one predicted by photoionization models (e.g., \citealt{GoadEtAl1993}; \citealt{KoristaEtAl2000, KoristaEtAl2004}). Moreover, \citet{OBrienEtAl1995} utilized the multicloud BLR models to confirm the non-linear response of the $\mgii$ line, resulting in the continuum-level dependent response function. This is well verified by our results that the emission-line responsivity is anti-correlated with Eddington ratio (i.e., high/low states).

Furthermore, our results have important consequences for future studies of $\mgii$ RM and SMBH mass estimation for a large set of 0.3 $< z < 2$ quasars. The statistically derived average value of $\rm{dlog}$$F_{line}$/$\rm{dlog}$$F_{cont} \approx$ 0.464, suggests a weak $\mgii$ BEL responsivity to continuum variations. This indicates that minimizing spectrophotometric errors is essential to revealing the intrinsic variability of $\mgii$. Only high signal-to-noise flux of $\mgii$ in monitoring campaigns of long duration can we expect to determine reliable lags between the $\mgii$ and 3000 \AA \ continuum. 

\section{Conclusions} \label{sec:con}

In this paper, we have statistically determined the relation between the magnitude differences of broad $\mgii$ and those of the 3000 \AA \ continuum based on the broad-line quasars taken from SDSS-I/II/III surveys, using a sample of 68 intermediate-redshift quasars, each with multiple ($\geq$ 2) observations and at least one S/N $\geq$ 5 narrow-line. The main conclusions are the following.

\begin{enumerate}
\item We found that $\mgii$ and the continuum variations are significantly correlated (Spearman $\rho = 0.593$). This is consistent with the idea that $\mgii$ varies in response to the continuum emission changes (see Figures \ref{f04}; Section \ref{subsec:vcc}).
\item Using the modified weighted least squares regress method, we confirmed that the slope between the variations in broad $\mgii$ emission-line and in the continuum (i.e., the responsivity $\alpha$) is not constant. But instead, we found that the slope anti-correlates with the quasar luminosity and/or Eddington ratio (i.e., high/low states; see Table \ref{tb2}; Figure \ref{f08}; Section \ref{subsec:sbc}).
\item We demonstrated the responsivity of $\mgii$ with an average $\bar{\alpha} \approx$ 0.464 (median 0.428) in our quasars, suggesting that high signal-to-noise flux measurements are statistically required to robustly detect the intrinsic variability and the time lag of $\mgii$ line (see Sections \ref{subsec:sbc} and \ref{sec:dis}).
\end{enumerate}

Generally speaking, a small slope would require high precision of the line flux measurements (i.e., small spectrophotometric errors and high S/N spectra) in order to obtain the time delay of $\mgii$ with respect to the ionizing continuum. One can also use the slope to constrain the physical parameters of the BLR. For instance, the optical thin gas would imply a negative response (see, e.g., \citealt{GoadEtAl1993}; \citealt{OBrienEtAl1994}). From this point, our results provide a useful diagnostic of physical conditions in the BLR for the sources with different luminosities.

\acknowledgments

We would like to thank the anonymous referee for constructive comments that led to an improved presentation. We are grateful to Chenwei Yang for giving friendly advice on our quasar sample selection and to Yue Shen for providing the UV and optical $\feii$ templates used in this work. We want to thank Chenwei Yang, Yue Shen, Qingfeng Zhu, Xiaobo Dong, Tuo Ji, Jianhui Lian and Junxian Wang for their useful help. This work was supported by the National Basic Research Program of China (Grant No. 2015CB857005), the National Natural Science Foundation of China (Grant Nos. 11233002 and 11421303), and the NSFC-CAS Joint Fund (Grant No. U1431229). M.Y.S. acknowledges support from the China Postdoctoral Science Foundation (Grant No. 2016M600485) and the National Natural Science Foundation of China (Grant No. 11603022).   
 
Funding for the SDSS and SDSS-II has been provided by the Alfred P. Sloan Foundation, the Participating Institutions, the National Science Foundation, the U.S. Department of Energy, the National Aeronautics and Space Administration, the Japanese Monbukagakusho, the Max Planck Society, and the Higher Education Funding Council for England. The SDSS Web Site is {\tt\string http://www.sdss.org/}. Funding for SDSS-III has been provided by the Alfred P. Sloan Foundation, the Participating Institutions, the National Science Foundation, and the U.S. Department of Energy Office of Science. The SDSS-III web site is {\tt\string http://www.sdss3.org/}.

SDSS-III is managed by the Astrophysical Research
Consortium for the Participating Institutions of the
SDSS-III Collaboration including the University of Arizona, the Brazilian Participation Group, Brookhaven
National Laboratory, Carnegie Mellon University, University of Florida, the French Participation Group,
the German Participation Group, Harvard University,
the Instituto de Astrofisica de Canarias, the Michigan
State/Notre Dame/JINA Participation Group, Johns
Hopkins University, Lawrence Berkeley National Laboratory, Max Planck Institute for Astrophysics, Max Planck Institute for Extraterrestrial Physics, New Mexico State University, New York University, Ohio State University, Pennsylvania State University, University of Portsmouth, Princeton University, the Spanish Participation Group, University of Tokyo, University of Utah, Vanderbilt University, University of Virginia, University of Washington, and Yale University.

\newpage

\bibliography{agn}




\end{document}